\documentclass[fleqn,12pt]{wlscirep} %andrey put 12, was 10
\usepackage[utf8]{inputenc}
\usepackage[T1]{fontenc}
\usepackage{color} %@Andrey
\usepackage{graphicx}% Include figure files
\usepackage{dcolumn}% Align table columns on decimal point
\usepackage{bm}% bold math

\DeclareMathOperator\erf{erf} %@Andrey
\title{On the interpretation of manifold ultrafast dynamics in supported graphene}

\author[1,+]{Rustam Gatamov}
\author[1,+,*]{Andrey Baydin}
\author[1]{Halina Krzyzanowska}
\author[1]{Norman Tolk}
\affil[1]{Department of Physics and Astronomy, Vanderbilt University, Nashville, TN 37235, USA}
\affil[*]{andrey.baydin@vanderbilt.edu}
\affil[+]{these authors contributed equally to this work}

\begin{abstract}
Understanding the ultrafast carrier dynamics of graphene on a substrate is a fundamental step in the development of graphene based opto-electronic devices.  Here, we present ultrafast pump-probe measurements of supported graphene on quartz for a range of pump fluences that enable us to observe both decreased and enhanced probe transmission regimes on a femtosecond timescale. Unexpectedly, at an intermediate pump fluence, an order of magnitude decrease in the relaxation time constant of the differential transmission is observed. By employing a number of different models to interpret our experimental data, we demonstrate the importance of considering both intra- and inter-band contributions to the dynamical optical conductivity in order to extract a more physical relaxation time constant of hot optical phonons.
\end{abstract}

\begin{document}

\flushbottom
\maketitle
\thispagestyle{empty}

\section*{Introduction}

Graphene is a single layer material of carbon atoms with high carrier mobility and unique opto-electronic properties which make it useful for a wide range of applications \cite{RevModPhys.81.109, NatPhot.4.611, RevModPhys.82.2673, Science.334.6056, doi:10.1021/nn901703e, 10.1038/nature10067}. The use of the advantageous properties of graphene in application to opto-electronic devices inevitably entails the generation of hot carriers with energies significantly exceeding the Fermi energy \cite{Science.334.6056, RevModPhys.81.109, NatPhot.4.611}. Therefore, carrier relaxation dynamics plays a central role in many proposed graphene based devices \cite{doi:10.1021/nl8033812, doi:10.1063/1.3413959, 4362685, PhysRevB.80.085109}. Especially with the scale of devices continuing to shrink, power capabilities being pushed to the limit and efficient heat removal becoming an issue it is hard to overestimate the importance of a more comprehensive understanding of carrier relaxation dynamics in graphene. 

There is an enormous ongoing effort by many groups to study graphene related relaxation dynamics using various techniques such as photocurrent measurements, ultrafast pump-probe spectroscopy, time resolved Raman spectroscopy, etc. \cite{doi:10.1021/nl8029493, PhysRevB.81.165405, PhysRevB.83.153410}. Ultrafast pump-probe spectroscopy has been particularly fruitful in providing valuable insights into electron-electron, electron-phonon and phonon-phonon interactions \cite{PhysRevLett.101.157402, Newson:09, PhysRevB.83.153410, Ruzicka:12, doi:10.1063/1.4799377, doi:10.1063/1.3291615, doi:10.1063/1.4982738, Malard2013}. Typically, after electrons and holes are excited into a non-thermal distribution by an ultrafast laser pulse, they thermalize into a Fermi-Dirac distribution through Coulomb interactions in tens of femtoseconds \cite{PhysRevLett.111.027403}. The cooling of the hot thermal population of carriers occurs through the emission of optical phonons. At the point when the temperatures of the electron and phonon systems equilibrate, the hot phonon bottleneck occurs, which significantly lessens the cooling rate \cite{doi:10.1063/1.3291615, doi:10.1063/1.4940902, doi:10.1021/nl201397a}. Subsequent cooling primarily results from the hot optical phonons undergoing anharmonic decay into acoustic phonons. However, in the case of supported graphene, direct coupling of the charge carriers to surface phonons in the polar substrates is a possible cooling channel \cite{PhysRevB.86.045413, PhysRevB.87.115432, PhysRevB.90.245436}. As predicted by theory and measured by experiments, the time constant of the hot optical phonon decay in graphene is of the order of a few picoseconds \cite{doi:10.1063/1.4940902, doi:10.1063/1.3291615, PhysRevLett.99.176802, PhysRevB.81.165405, Malard2013, doi:10.1063/1.4982738, doi:10.1063/1.4799377}. The hot phonon bottleneck bears far reaching implications for device performance, in particular the timescale of the photoresponse in opto-electronic devices \cite{Science.334.6056, PhysRevLett.95.236802, PhysRevLett.95.155505, PhysRevB.92.075414}. Hence, full understanding of the cooling pathways for excited carrier and hot phonons are necessary for device applications. %andrey: In the previous ultrafast pump-probe studies of the hot phonon dynamics in graphene as a function of pump fluence, the behavior of the phenomenological relaxation time constant of transmission was found to be nearly monotonic \cite{doi:10.1063/1.4799377, doi:10.1021/nl201397a, HUANG20111657}. %Here, we report the pump probe measurements that reveal the strong nonlinear behavior of %the hot phonon 

Here, we study ultrafast carrier and phonon dynamics in CVD graphene on a quartz substrate using pump-probe spectroscopy in transmission geometry. Unlike other numerous reports on graphene on quartz, we cover a range of pump fluences that allows us to observe both positive and negative transmission changes following photo-excitation, and, most importantly, an enhanced decay of differential transmission  arising from the competing contributions of intra- and inter-band transitions at intermediate pump fluences. In addition, we show that if the experimental data is analyzed phenomenologically by fitting it to a bi-exponential decay as opposed to a more physics based model, it may lead to an incorrect interpretation as discussed in more detail in the Results section. 

%andrey we report measurements of the strong nonlinearity of the relaxation time constant with respect to the density of photo-excited carriers as obtained from the ultrafast pump-probe measurements in the transmission geometry of the CVD graphene on a quartz substrate. %around a certain pump fluence, which, to the best of our knowledge, has not been observed before and is not predicted by the electron-phonon model.

\section*{Results}
\subsection*{Experimental data}
\begin{figure}[ht!]
    \centering
    \includegraphics{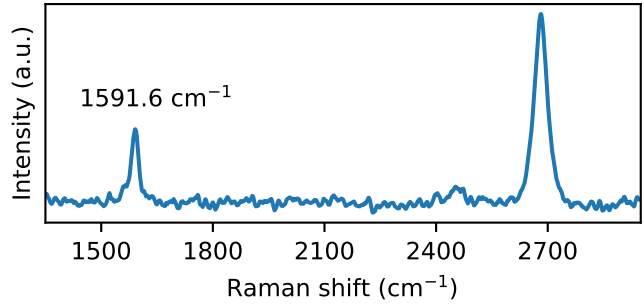}
    \caption{Raman spectrum of graphene on quartz taken with 532 nm CW laser.}
    \label{fig:raman}
\end{figure}

Single layer CVD graphene (Graphene Supermarket) was transferred onto a quartz substrate by a wet transfer method (see Methods). The quality and uniformity of the transferred graphene layer was confirmed by Raman spectroscopy using a Thermo Scientific DXR Raman Microscope. The measured Raman spectrum of graphene on quartz is shown in Fig. \ref{fig:raman}. The Fermi level was estimated to be 316 meV from the 11.6 cm$^{-1}$ shift of the G peak position \cite{PhysRevB.91.205413}. Ultrafast pump-probe transmission measurements were carried out using a Ti:Sapphire oscillator producing 120 fs pulses at a 76 MHz repetition rate. Pump and probe wavelengths were both set to 800 nm (1.55 eV). Both beams were focused onto the specimen with spot diameters of 80 $\mu$m and 50 $\mu$m for pump and probe, respectively. The pump beam was chopped using a SR540 optical chopper operating at about 4kHz. The differential transmission signal was detected using a Lock-In amplifier. 

\begin{figure}
    \centering
    \includegraphics{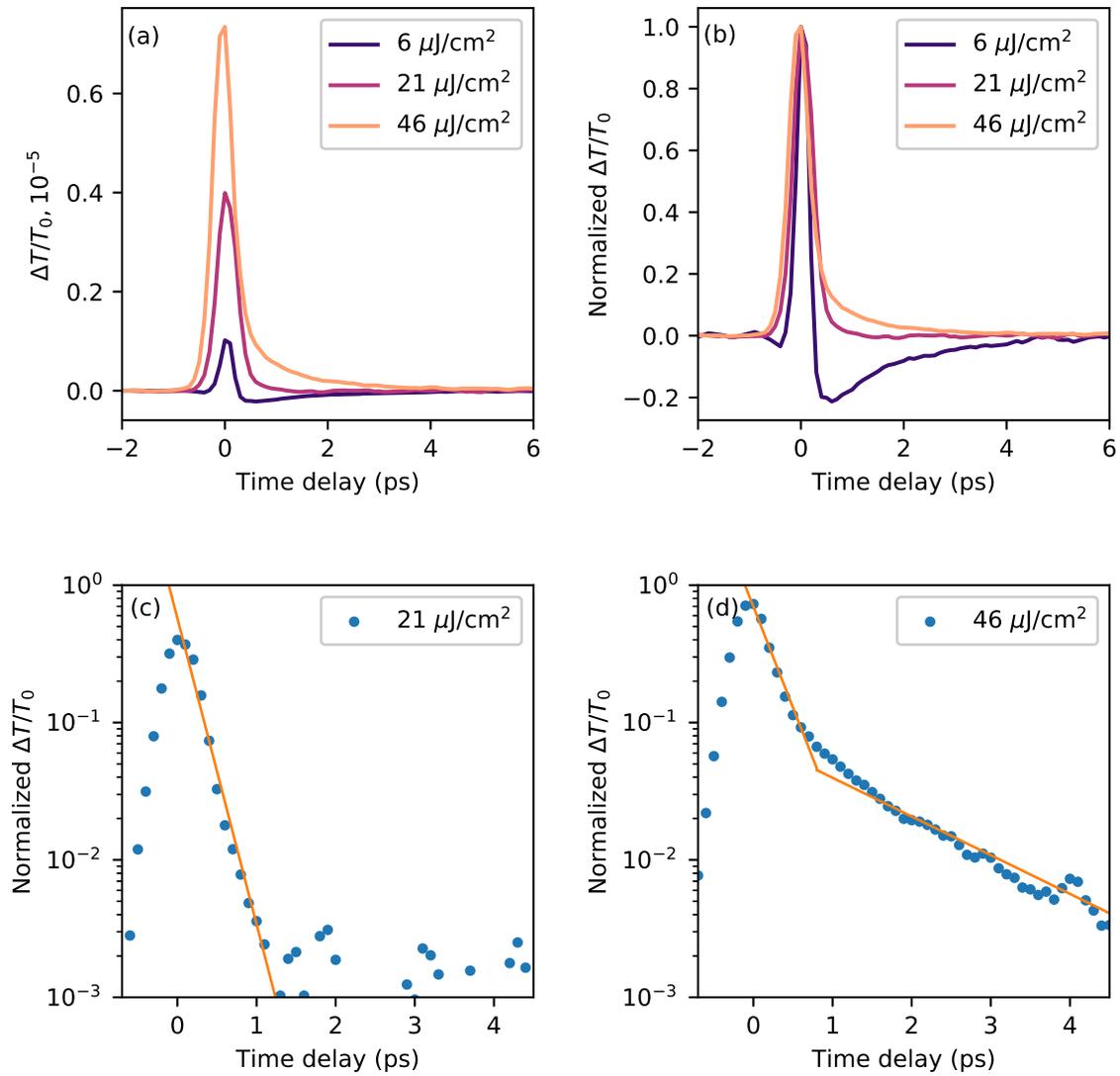}
    \caption{Pump probe response of graphene on quartz for different pump fluences}
    \label{fig:transmission}
\end{figure}

Figure \ref{fig:transmission}a shows differential transmission data taken at three different pump fluences. After the initial photoexcitation, the transmission dynamics can be either fully positive or first positive and then negative depending on the pump fluence. At high fluences (above 33 $\mu$J/cm$^2$), the change in transmission is positive at all times, while for lower fluences (below 18 $\mu$J/cm$^2$) the initial positive change in the transmission becomes negative past $\approx$0.5 ps. 

Electron and phonon dynamics in graphene on quartz has been extensively studied at different pump fluences \cite{doi:10.1021/nl201397a, HUANG20111657, doi:10.1063/1.4799377}. However, there has not been a previous study showing what happens at a particular pump fluence when the change in transmission past $\approx$0.5 ps crosses over from negative to positive values (see data at the fluence of 21 $\mu$m/cm$^2$ in Fig. \ref{fig:transmission}a,b). Interestingly, at at this fluence, the differential transmission decay does not have a slow component as opposed to the lower and higher fluence cases. This behaviour is seen more clearly on the log scale in Fig. \ref{fig:transmission}c,d.

As the experimental data has fast and slow decay components, it can be fitted to a bi-exponential decay convoluted with Gaussian pump and probe pulses using the following equation\cite{prasankumar2016optical}: 
\begin{equation}
\begin{split}
    D_{1}\exp\left(-\frac{t-t_{0}}{\tau_{1}}\right)\left[1-\erf\left(\frac{w}{2\tau_{1}}-\frac{t-t_{0}}{w}\right)\right]+D_{2}\exp\left(-\frac{t-t_{0}}{\tau_{2}}\right)\left[1-\erf\left(\frac{w}{2\tau_{2}}-\frac{t-t_{0}}{w}\right)\right],
\end{split}
\label{eq:bi-exp-fit}
\end{equation}
where $D_1, D_2, t_0, \tau_1, \tau_2$ are fitting parameters. $\omega=\sigma\sqrt{2}$ is the width of the autocorrelation function with $\sigma$ being the width of the the Gaussian pump and probe pulses. The time constants $\tau_1$ and $\tau_2$ refer to fast and slow decay processes, respectively. It is important to note, that the fast decay characterized by $\tau_1$ is usually attributed to thermalization between electrons and optical phonons while the slow decay characterized by $\tau_2$ is often assigned to the relaxation of hot optical phonons in graphene\cite{Dawlaty2008, Huang2010, HUANG2011, Shang2011, Gao2011,winnerl2013time,Oum2014}. Since our pulse width is 120 fs, we cannot accurately measure the time constant $\tau_1$. Therefore, we report and focus only on the slow decay characterized by the time constant $\tau_2$.

The fitting results using equation (\ref{eq:bi-exp-fit}) are shown in figure \ref{fig:tau2}a. Surprisingly, the $\tau_2$ dependence on pump fluence is not monotonic and drops by an order of magnitude at intermediate fluences. As mentioned above, the slow decay characterized by $\tau_2$ is often considered as the relaxation time constant of hot optical phonons. At first glance, one could conjecture that the mitigation of the hot phonon bottleneck is observed at the pump fluence of about $21 \mu$J/cm$^2$. However, such a conclusion is shown to be incorrect. Consequently, in order to better understand our data, we have utilized a model that is more physics based and appropriately modified for our system. 

\subsection*{Physical model}

\begin{figure}
    \centering
    \includegraphics{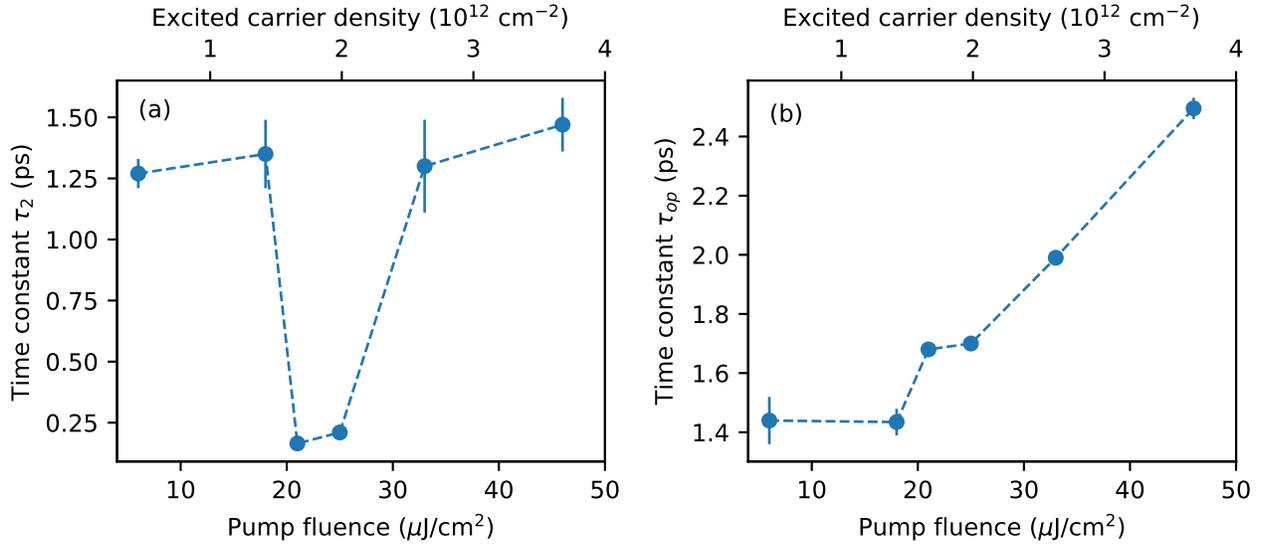}
    \caption{Relaxation time $\tau_2$ related to hot phonon decay as obtained from a) bi-exponential fit of differential transmission data and b) the physical model that accounts for inter- and intra-band transitions.\textbf{}}
    \label{fig:tau2}
\end{figure}

Transmission of the single layer graphene on a substrate is given by\cite{Choi2009}
\begin{equation}
    t(\omega) = \frac{1}{1+\sigma(\omega)Z_{0}/(1+n_{s})},
    \label{eq:transmission}
\end{equation}
where $\sigma(\omega)$ is the optical conductivity of graphene, $Z_{0}$ is the vacuum impedance and $n_{s}$ is the refraction index of the substrate.

The optical conductivity of single layer graphene is given by\cite{Dani2012}
\begin{equation}
\begin{split}
    \frac{\text{Re}\left(\sigma\left(\omega\right)\right)}{\sigma_{Q}}=\frac{4k_{B}T}{\pi\hbar}\left[\ln\left(1+e^{E_{F}^{e}/k_{B}T}\right)+\ln\left(1+e^{E_{F}^{h}/k_{B}T}\right)\right]\frac{\tau}{\omega^{2}\tau^{2}+1}+\\ \frac{1}{2}\left[\tanh\left(\frac{\hbar\omega-2E_{F}^{e}}{4k_{B}T}\right)+\tanh\left(\frac{\hbar\omega-2E_{F}^{h}}{4k_{B}T}\right)\right],
\end{split}
\label{eq:conductivity}
\end{equation}
where $\sigma_{Q}=e^2/4\hbar$ is the universal quantum conductivity, $k_B$ is Boltzmann constant, $T$ is carrier temperature, $E_F^e (E_F^h)$ is the electron (hole) Fermi energy, $\tau$ is the momentum scattering time. The first term in equation (\ref{eq:conductivity}) describes a Drude-like intraband contribution to optical conductivity whereas the second term arises from interband transitions. The measured differential transmission is proportional to $-\Delta\left[\text{Re}\left(\sigma\left(\omega\right)\right)/\sigma_{Q}\right]$ according to equation (\ref{eq:transmission}). Considering high n-type doping ($E_F^e>>2k_B T$) and neglecting the contribution from minority carriers (holes), the changes in the optical conductivity can be written as
\begin{equation}
    -\Delta\left[\frac{Re\left(\sigma\left(\omega\right)\right)}{\sigma_{Q}}\right]=\frac{4E_{F}^{e}}{\pi\hbar}\frac{\tau}{\omega^{2}\tau^{2}+1}+\frac{1}{2}\tanh\left(\frac{\hbar\omega-2E_{F}^{e}}{4k_{B}T}\right). 
    \label{eq:deltacond}
\end{equation}
The Fermi energy, $E_F^e$, and temperature, $T$, of electrons in the equation (\ref{eq:deltacond}) are, in general, time dependent quantities.

The dynamics of electron temperature, $T$ is modeled by bi-exponential decay (see Methods):
\begin{equation}
    T(t)=T_{0}+\frac{(T_{1}-T_{0})}{2}\left[(1-k)e^{-t/\tau_{e-ph}}+(1+k)e^{-t/\tau_{op}}\right],
    \label{eq:temp-bi-exp}
\end{equation}
where $T_0$ is the equilibrium lattice temperature, $T_1$ is the temperature upon photo-excitation, $k, \tau_{e-ph}, \tau_{op}$ serve as fitting parameters. For graphene, the heat capacity of phonons is much larger than the heat capacity of electrons. This indicates that $\tau_{e-ph}$ can be thought of as the thermalization time of electron and phonons systems, while $\tau_{op}$ is the hot phonon relaxation time (see Methods). %Note that $\tau_{op}$ is expected to change with excited carrier densities as it represents characteristic relaxation time and not the optical phonon lifetime. 

The initial electron temperature, $T_1$, and Fermi energy, $E_F^e$ are estimated from energy conservation to be\cite{Choi2009,Dani2012}:
\begin{equation}
    U_e(T_0,N)+\Delta Q=U_e(T_1, N+n)+U_h(T_1, p),
\end{equation}
where
\begin{equation}
    U_{e,h}=\frac{4k_{B}^{3}T^{3}}{\pi v^{2}\hbar^{2}}F_{2}\left(\frac{E_{F}^{e,h}}{k_{B}T}\right),
\end{equation}
\begin{equation}
    n,p=\frac{2k_{B}^{2}T^{2}}{\pi v^{2}\hbar^{2}}F_{1}\left(\frac{E_{F}^{e,h}}{k_{B}T}\right).
    \label{eq:carrier-density}
\end{equation}
$F_1$ and $F_2$ are the first- and second-order Fermi integrals, respectively. $v = 10^6$ m/s is the Fermi velocity in graphene, $N$ is the electron concentration at equilibrium. The time dynamics of the Fermi energy, $E_F^e$ is taken into account by solving equation (\ref{eq:carrier-density}) with carrier densities recombining exponentially, $n=N+n_0 e^{-t/\tau_R}$. The effective recombination time $\tau_R =1.3$ ps\cite{Choi2009} was fixed for all fits. 

Figure \ref{fig:tau2}b shows the pump fluence dependence of the relaxation time constant $\tau_{op}$. Clearly, $\tau_{op}$ increases with temperature and does not have a dip at intermediate pump fluences as in the case of $\tau_2$. The changes in $\tau_{op}$ as a function of temperature are not surprising as we have ignored the temperature dependence of the electron-phonon coupling constant and electron and phonon heat capacities. For a more rigorous modelling and extraction of more correct optical phonon relaxation times, rate equations including phonon generation/recombination rates can be employed\cite{doi:10.1063/1.3291615}. However, our approach is simpler to implement and computationally less expensive. The comparison between the fluence dependence of $\tau_2$ defined in equation (\ref{eq:bi-exp-fit}) and $\tau_{op}$ defined in equation (\ref{eq:temp-bi-exp}) illustrates the importance of having chosen a more physical model which may be generally useful in the interpretation of ultrafast pump-probe data in supported graphene systems.

\section*{Discussion}
%andrey: due to increased contribution from intra-band transitions \cite{Kadi2014}. In general, the sign of the differential transmission can be explained by the interplay between inter-band and intra-band electron transitions \cite{Malard2013}. 

We have investigated ultrafast carrier-phonon relaxation in  supported graphene using transmission pump-probe spectroscopy in the range of pump fluences that show distinctly different regimes in the relaxation dynamics. One of the main results is the observation of an order of magnitude faster decay of the differential transmission at intermediate pump fluences as compared to low and high fluences. Such unconventional behavior of the relaxation is due to the competitive interplay between inter- and intra-band transitions as they contribute to the differential optical conductivity with opposite signs. Indeed, our model takes into account both intra- and inter-band transitions and is a modified version of a model typically used in many research efforts\cite{Choi2009, Dani2012, Malard2013, CHEN2016}. The optical phonon time constant, $\tau_{op}$, extracted from fitting the physical model to our data increases with pump fluence. In contrast to this finding, the time constant, $\tau_2$, obtained from a phenomenological bi-exponential fitting of the transient transmission data, exhibits an entirely different trend. It is important to note that $\tau_{2}$ is often attributed to optical phonon relaxation, which will result in a misleading conclusion (see Fig. \ref{fig:tau2}a). For low fluence, the observed dynamics is governed by intra-band transitions, while for high fluence it is governed by inter-band transitions. However, our results show that it is crucial to be aware of the critical fluence for a particular sample when both intra- and inter-band transitions contribute. Particularly, inter- and intra-band transitions should be accounted for when comparing graphene dynamics on different substrates where graphene can have a different Fermi level depending on a substrate. In conclusion, this work adds one more significant factor in the characterization of the diversity of ultrafast dynamics in graphene and is of significance to graphene related research and applications. 

%they contribute to the differential optical conductivity with opposite signs \cite{Tomadin2018, Malard2013, Chen2014, Kadi2014}

\section*{Methods}

\subsection*{Sample preparation}
Single layer CVD graphene on a copper foil was obtained from Graphene Supermarket. Graphene was transferred onto a quartz substrate using the wet transfer method. First, poly-methyl methacrylate (PMMA) was spun coated on top of the graphene/copper sample at 4500 rpm. Then, the sample was left in Iron Chloride (FeCl$_3$) for 24 hours to etch away copper. After the copper foil is etched, the PMMA/graphene layer was cleaned of the copper etchant residue by consequently dipping it in Deionized Water (DI) and then leaving it in DI for 24 hours. Next, the PMMA/graphene sample was mechanically transferred onto a quartz substrate. As the last step, we removed PMMA layer by dissolving it in acetone, which followed by an isopropanol (IPA) rinse. 

\subsection*{Ultrafast pump-probe spectroscopy}
Ultrafast pump-probe spectroscopy measurements were carried out in transmission geometry using Ti:Sapphire Coherent Mira 900 oscillator producing 120 fs pulses at 76 MHz repetition rate. Pump and probe wavelengths were set to 800 nm (1.55 eV). Both beams were focused onto the specimen with spot diameters of 80 $\mu$m and 50 $\mu$m for pump and probe, respectively. Pump and probe beams were cross-polarized for improved detection. The pump beam was chopped using a SR540 optical chopper operating at 4kHz, which served as a reference the lock-in amplifier. 

\subsection*{Model of electron temperature}

Upon photo-excitation, systems of electrons and optical phonons obey the rate equations, where the changes in the temperature of acoustic phonons is neglected. 
\begin{equation}
    \begin{cases}
    \dot{T} = \alpha(T_{op} - T),\qquad\qquad\qquad\qquad T(0)=T_1,\\
    \dot{T}_{op} = \beta(T-T_{op}) + \gamma(T_{0}-T_{op}), \qquad T_{op}(0) = T_{0},\\
    \dot{T}_{ac} = 0,\qquad\qquad\qquad\qquad\qquad\qquad T_{ac}(0) = T_{0},
    \end{cases}
    \label{eq:temperaturesystem}
\end{equation}
where $\alpha=G_{el,op}/C_{el}$, $\beta=G_{el,op}/C_{op}$, $\gamma=G_{op,ac}/C_{op}$ with $G$ and $C$ denoting coupling constants and heat capacities of electrons and phonons. Assuming the coefficients are constant with respect to temperature, this system of differential equations (\ref{eq:temperaturesystem}) can be solved analytically. The solution for the electron temperature is
\begin{equation}
    T(t)=T_{0}+\frac{(T_{1}-T_{0})}{2}\left[(1-k)e^{-t/\tau_{e-ph}}+(1+k)e^{-t/\tau_{op}}\right],
\end{equation}
where $k = (-\alpha+\beta+\gamma)/\sqrt{(-\alpha+\beta+\gamma)^2+4\alpha\beta}$, $\tau_{e-ph}=2/[(\alpha+\beta+\gamma)+\sqrt{(\alpha+\beta+\gamma)^2-4\alpha\gamma}]$ and $\tau_{op}=2/((\alpha+\beta+\gamma)-\sqrt{(\alpha+\beta+\gamma)^2-4\alpha\gamma})$. It is worth noting that since $\alpha>>\beta,\gamma$ (due to $C_{el} << C_{op}$), $\tau_{e-ph} \approx 1/\alpha$ and $\tau_{op} \approx 1/\gamma$. This indicates that $\tau_{e-ph}$ can be thought of as the thermalization time of electron and phonons systems, while $\tau_{op}$ is hot phonon relaxation time.

%Therefore, by making $\tau_{2}$ one of the fitting parameters we are able to approximately measure $1/\gamma$, which is referred to as the optical phonon scattering time constant. In general, heat capacity and electron-phonon coupling, and consequently, $\alpha, \beta, \gamma$ are functions of temperature. However, for graphene heat capacity of phonons is much larger than heat capacity of electrons. 

\section*{Acknowledgements}

The authors acknowledge the ARO for financial support under Contract No. W911NF-14-1-0290. Portions of this work were completed using the shared resources of the Vanderbilt Institute of Nanoscale Science and Engineering (VINSE) core laboratories.

\section*{Author contributions statement}

R.G. conducted the experiment, R.G. and A.B. analyzed the results, R.G. and A.B co-wrote the manuscript with input from H.K. and N.T. A.B. supervised the project. All authors discussed the results and reviewed the manuscript. 

\section*{Additional information}

The authors declare that they have no competing interests.

\bibliography{bibliography}
\end{document}